# LESSONS FROM THE COINSEMINAR

Peter Gloor, Maria Paasivaara, Christine Miller

MIT, Aalto University, Illinois Institute of Technology
Cambridge USA, Helsinki Finland, Chicago USA
pgloor@mit.edu, maria.paasivaara@aalto.fi, cmille31@stuart.iit.edu

## ABSTRACT

This paper describes lessons learned from teaching a distributed virtual course on COINs (Collaborative Innovation Networks) over the last 12 years at five different sites located in four different time zones.

## INTRODUCTION

For the last 12 years the authors have been involved in teaching a distributed seminar where student teams with participants from MIT, Savannah College of Art and Design, Illinois Institute of Technology, Aalto University, University of Cologne, and University of Bamberg collaborated as virtual distributed teams, tackling problems of social media analysis and other COINs-related issues. The course consists of an introductory block course taught on-site, followed by four months of virtual collaboration by distributed student teams. Over these 12 years we experimented with many different ways of increasing creativity and productivity of the student teams. We have collected student feedback by surveys and interviews.

## VIRTUAL MIRROR OF COMMUNICATION

One of the key concepts we are using to increase communication is "virtual mirroring". By looking at the communication behavior of the teams by mining their e-mail archive using Condor, we are able to track the health of the social life of a team (Gloor et. al 2012).

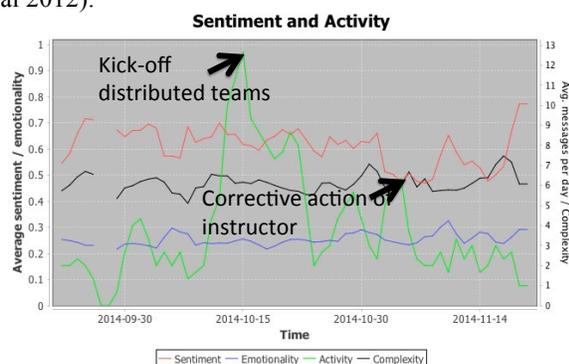

*Figure 1: Activity, sentiment, and emotionality Sept 21 to November 19, 2014 of COINs14 course*

Figures 1 and 2 illustrate sentiment and centralities over time of the fall 2014 COINS seminar with participants from Aalto, MIT, IIT, U. Cologne and U. Bamberg. The analysis is done by using the mailbox of the main instructor as a proxy for the organizational memory of the course. Note how the sentiment over time in figure 1 is going down. We have found that this is a sign of a healthy culture, as team members are changing from a supportive communication style, which is mostly positive, to a more honest language that mentions both positive and negative issues as they come up.

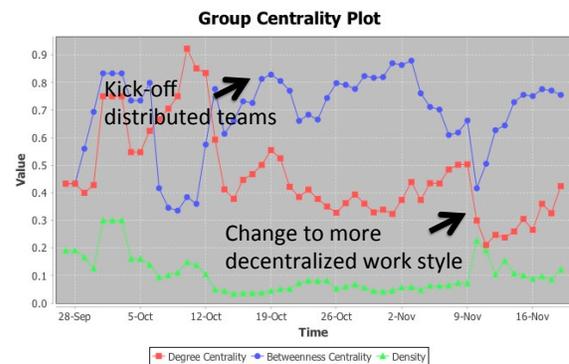

*Figure 2: group betweenness centrality, degree and density Sept 21 to November 19, 2014*

Figure 2 illustrates the drop in group betweenness centrality after November 12, with the mode of teamwork shifting from being dominated by the main instructor with high centrality to a mode where teams are working more independently.

## LESSONS FOR MANAGING DISTRIBUTED TEAMS

Looking at the development of the course over the last 12 years confirms that many of the seminal results of (Hackmann 2011) also apply in the virtual world. Hackmann specifies a 60-30-10 rule of how leaders (in our case the teachers) can influence performance of their teams: 60% of the leader-influenced performance of a team depends on preparation and prework of the leader (teacher), 30% of the team performance depends on the launch execution, and 10% depends on what the leader does once the team is underway. One of our students commented that during the COINS course he *"learned the importance of effective communication and the need of one person to take the lead position within the heterogeneous group to focus on productive outcomes."* Hackmann identifies six conditions that foster team effectiveness (Hackmann 2011):

1. **The team is a real team, with a bounded set of people**. We encourage virtual teams to spend enough team-building time using Web conferencing with the camera turned on, where members take time to get acquainted with each other. Students have commented that this is one of their major learning: *"Use time to get to know other team members in order to create shared understanding and knowledge about each other, this will help you work together later on."*
2. **The team's purpose is challenging and consequential, with desired end states clearly stated, but the means to execute largely left to the team**. We try to set clear project goals, such as for instanced finding new trends for Cystic Fibrosis in social media, but leave it up to the teams to self-organize and decide on tools and methods to accomplish their goals. Students have liked our challenging topics: *"Having a truly inspiring topic keeps you motivated"*
3. **The team has the right number of people – a slightly understaffed team is best**. We found that having teams composed of members from two to three locations, with ideally two members co-located at the same location works best. This leads to a total team size of three to six.
4. **The team has clear norms of conduct that promote full utilization of team members' capabilities.** Using virtual mirroring, and educating members about intercultural collaboration has become an integral part of the course. We have also encouraged students to share their team members what they can and would like to learn and how they could contribute to the project right in the beginning, as one of our students put it:*"First they need to know about their own strengths and their weaknesses, and they have to talk about it right before they want to start their project. […] We can actually share our feelings and our rationale about how we choose this project and how we think that we can contribute to this project."*
5. **The team has the resources it needs to accomplish the task.** Students have been trained in the initial block course in the tools and methods (Condor, COINs, Coolhunting and Coolfarming) that are applied in the course, which students have found important: *"It is much more convenient for people to come to a group project when they know, they have the same knowledge."* If needed, we also provide Web hosting space for the teams to develop cloud-based solutions.
6. **The team receives competent, well-timed coaching.** Each team gets a coach assigned who tries to respond to requests from teams within the same day, and also assists when students experience technical difficulties.

## LOOKING AHEAD: APPLICATIONS AND RELEVANCE OF THE COINSEMINAR

For students, the benefits derived from participation in the COINSEMINAR are multifold. For example, learning to successfully navigate within global virtual teams challenges students to articulate and demonstrate their respective disciplinary contributions in multicultural and multiple disciplinary settings. Client-based projects require that students establish ways of working together to define, research and develop solutions using both 'off-the-shelf' and proprietary software tools. Opportunities to apply network perspectives in research are increasing rapidly across many fields. The relevance of the seminar is supported by evidence in projects such as the *Chicago Model*, a three-dimensional digital model of the city, created by the Chicago Architecture Foundation (CAF) as a canvas for visualizing city data. The networked view of data flows within the city provide a demonstration of how "Big Data" that encompasses everything from data collected by environmental sensors *to messages on social media* is being used by designers, planners and citizens to understand and solve problems. Projects such as the *Chicago Model* inspire us to consider the coming challenges in harnessing and managing massive data flows between human networks and the machine social networks in the emerging Internet of Things (IoT) (daCosta 2013).